# Stability, Tunneling Characteristics and Thermoelectric Properties of TeSe$_2$ allotropes

Munish Sharma

*Department of Physics, School of Basic and Applied Sciences, Maharaja Agrasen University, Baddi 174103, India.*

(December 7, 2020)

*Corresponding Author

Email: munishsharmahpu@live.com

**Abstract**

The waste heat management becomes very important with increasing energy demand and limited fossil resources. Here, we demonstrate thermoelectric performance of allotropic TeSe$_2$. Based on the first-principle calculations, we confirm the energetic and kinetic stability of five TeSe$_2$ allotropes. We predict δ-TeSe$_2$ as a new direct band gap semiconductor having 1.60 eV direct band gap. All the TeSe$_2$ allotropes exhibit band gap in UV-Vis region. The structural phases are clearly distinguished using simulated scanning tunnel microscopy. The room temperature Seebeck coefficient is maximum of 4 V/K for δ-TeSe$_2$. We show that room temperature thermoelectric figure of merit (*ZT*) can reach up to 3.1 with p-type doping in δ-TeSe$_2$. Moreover, temperature and chemical potential tuning extends the thermoelectric performance of TeSe$_2$ allotropes. We strongly believe that our study is compelling from an experimental perspective and holds a key towards fabrication of thermoelectric devices based on TeSe$_2$.

# 1. Introduction

The discovery of graphene [1] and studies on its variety of allotropic forms [2, 3] has triggered research interest in new allotropic two dimensional (2D) materials to cover large spectrum of applications. The layered transition metal dichalcogenides (TMD's) have become popular due to their distinctive properties in allotropic forms for next generation thermoelectric devices [4-7]. In general, dimensionless figure of merit (*ZT*) typically quantify the efficiency of a thermoelectric material. Figure of merit, $ZT=S^2GT/(K_e+K_{pk})$ proportionally depend on the Seebeck coefficient (*S*), electrical conductance (*G*), electronic thermal conductance ($K_e$), phonon thermal conductance ($K_{ph}$). The optimization of thermoelectric efficiency in a thermoelectric material is key issue as all these thermoelectric parameters are interdependent.

The maximization of power factor ($S^2GT$) involve tuning of electronic properties and minimizations of lattice thermal conductance require phonon engineering. Alloying, doping, atomic defects, applied strain are reported effective strategies to improve ZT [6, 8-12]. However, the quantum confinement effects play an important role to enhance power factor and reduce thermal conductivity when material is scaled to nano-size [13, 14]. In the past few years, mono-elemental monolayers or few layers from group-III (e.g. borophene), group-IV (e.g. silicene, stanene) and group-V (e.g. phosphorene, antimonene) have been synthesized [15-17]. There is lack of reports on the group-VI based chalcogenide monolayers until the tellurene has been successfully synthesized on graphite substrate [18, 19].

The chalcogenide monolayers have attracted much attention as they exhibit tunable electronic band gap, unique non liner optical response [20], large piezoelectricity, high carrier mobility (103 cm$^2$V$^{-1}$s$^{-1}$), low lattice thermal conductivity [21] and excellent thermoelectric efficiency [22]. It is notable that few layers of tellurene has shown a unique polytypism during its synthesis using Physical Vapor Deposion (PVD) and Pulsed laser deposition (PLD) technique [23]. The polytypes in 2D tellurene has been further confirmed by temperature dependent Raman studies. A structural phase transitions in few layers tellurene has been observed under charge doping. [24] and electric field [25]. The chalcogenide monolayers based janus structures and binary forms are being explored by considering the structural similarities [26-28]. For example, α-TeSe$_2$ shares the similar structure as tellurene [29]. It has been reported that α-TeSe$_2$ exhibit tunable electronic and mechanical properties, optical anisotropy [27, 29], large piezoelectricity [28] and high carrier mobility [30] as compared to their elemental forms. Liu *at.al.* [26] have reported lower thermal expansion coefficient in TeSe$_2$ monolayer as compared to Selenene. Motivated by experimental findings and theoretical investigations on polymorphic chalcogenide monolayers we report structural stability, electronic and thermoelectric properties of six TeSe$_2$ monolayer allotropes.

## 2. Simulation Details

The spin polarized calculations have been performed using DFT based SIESTA code [31] for our calculations. We have used well tested, norm conserving, relativistic Troullier Martin pseudo potential [32] in fully separable Kleinman and Bylander forms. A Generalized gradient approximation within PBE parameterization has been employed to describe Exchange correlation potential. A vacuum buffer space of ~20 Å has been set along z-axis to minimize pseudo interaction between periodic images to model two dimensional systems. A double zeta polarized numerical atomic orbitals (NAOs) basis set have been used throughout the geometry optimization with confinement energy of 0.01 Ry. A standard conjugate-gradients (CGs) technique has been used to get total energy minima. A plane wave cutoff energy of 450 Ry has been used to calculate charge densities and DFT potentials. The lattice vectors and atomic positions were fully optimized using standard conjugate gradient technique. The atomic forces have been optimized until the forces on atoms are smaller than 0.02 eV/Å. The Brillouin zone have been sampled using a dense Monckhorst-Pack mesh of 15x15x1 for α, γ, σ polymorph and 15x9x1 for β, δ, η polymorphs. Phonon calculations has been carried out using supercell and finite displacement approach as implemented in Phonopy code [33]. A 4x4x1 supercell has been used to get the phonon thermal conductance ($K_{ph}$). The mean-field Hamiltonian obtained from the converged DFT calculations has been combined with non-equilibrium Green's function (NEGF) method to obtain transmission coefficient T(E) of the electrons with energy E passing from left to right electrode in GOLLUM code [34]. The thermoelectric properties such as electrical conductance G(T), the electronic thermal conductance $K_e$(T) and the Seebeck Coefficient S(T) can be calculated from the transmission probability T(E) as [34]

$$G(T) = \frac{2e^2}{h} L_0$$

$$S(T) = -\frac{L_1}{eTL_0}$$

$$K_e(T) = \frac{L_0 L_2 - L_1^2}{hTL_0}$$

$$L_n(T) = \int_{-\infty}^{+\infty} dE (E - E_F)^n T(E) \left( \frac{-\partial f(E)}{\partial (E)} \right)$$

Where, $E_F$ is Fermi energy, $f(E)$ is the Fermi–Dirac probability distribution function, $T$ is temperature, $e$ is electron charge and $h$ is the Planck's constant.

## 3. Results and Discussion
### 3.1 Structural Stability

Throughout this work we have considered six different allotropes of TeSe$_2$ which have been denoted with Greek letters as α, β, γ, δ, η and σ. The atomic structure of two dimensional TeSe$_2$ polymorphs has been shown in figure 1 and optimized structural parameters and structural anisotropy have been tabulated in table 1. The α, β, γ, σ allotrope contains three atoms per unit cell and δ, η allotrope contains six atoms per unit cell. Most of our calculated values for known α, β allotrope are in very close agreement with the available values in the literature [27, 30, 35]. For example, for α-TeSe$_2$, we find lattice constant of 4.03 Å, which is in close agreement with value of 3.98 Å reported by Naseri *et. al.* [27] and 3.99 Å reported by Debela *et. al.* [35]. The atomic arrangements in different allotropes lead to change in thickness of TeSe$_2$ monolayer. The TeSe$_2$ layer thickness has been found to be maximum of 3.78 Å for η allotrope and minimum of 1.66 Å for σ allotrope. The layer thickness is generally less than that observed in tellurene allotropes [36]. Note that α, γ allotrope exhibit isotropic nature while β, δ, η and σ exhibit anisotropic nature. The structural anisotropy is maximum in δ phase and minimum in β phase.

**Table 1** The optimized lattice parameters (*a, b*), thickness (*d*), structural anisotropy (*K*) of TeSe$_2$ allotropes using GGA-PBE functional. N represents the number of atoms in unit cell. Structural anisotropy has been calculated as *K=|(a-b)/(a+b)|*. Other reported values are also given.

| Phase | N | a | b | d | K |
|---|---|---|---|---|---|
| α | 3 | 4.03, 3.98[a], 3.99[b] | 4.03, 3.98[a], 3.99[b] | 3.37, 3.16[a] | - |
| β | 3 | 4.10, 3.86[c] | 5.29, 5.08[c] | 1.92 | 0.12 |
| γ | 3 | 3.88 | 3.88 | 3.73 | - |
| δ | 6 | 9.18 | 4.13 | 2.52 | 0.38 |
| η | 6 | 8.06 | 4.02 | 3.78 | 0.33 |
| σ | 3 | 4.04 | 5.28 | 1.66 | 0.13 |

[a]Ref [27], [b]Ref [35], [c]Ref [30]

In order to ensure the energetic stability of the considered 2D allotropes of TeSe$_2$ we have calculated the relative formation energy ($\Delta E_{Rel}$) of all polymorphs as compared to α-phase which is reported as stable phase in literature [26, 27]. The relative formation energy has been calculated as $\Delta E_{Rel} = E_p - E_\alpha$, where $E_p$ ($p$ = γ, β, δ, η and σ) and $E_\alpha$ are total energy per atom for respective allotrope. All the considered allotropes have $\Delta E_{Rel}$ less than 30 *meV* (figure 1), indicating their energetic stability and ease of formation. The α-phase is found to be energetically more favorable which is consistent with the earlier report [26, 27]. However, it is noticeable that except γ-TeSe$_2$ the difference in $\Delta E_{Rel}$ among all phases are less than 5 *meV*. A very narrow range of $\Delta E_{Rel}$ for different allotropes indicate the possibility of phase coexistence in experiments.

Furthermore, the dynamical stability of different allotropes have been confirmed by phonon calculations. Figure 2 represent the phonon dispersion of six TeSe$_2$ allotropes. The Phonon frequencies are found to be positive for all considered allotropes except for γ allotrope indicating kinetic stability of five allotropes and kinetic instability of γ allotrope in free standing state. The γ allotrope have out-of-plane acoustical modes with imaginary frequencies at high symmetry K-point on Brillouin zone indicating its stabilization on suitable substrate which can minimize the out-of-plane vibrations.

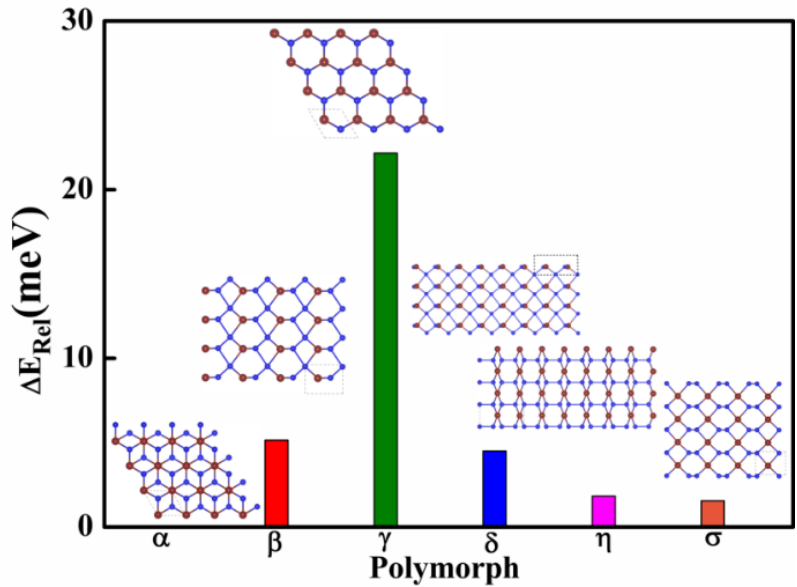

**Figure 1.** Relative formation energy per atom for TeSe$_2$ allotropes. Brown (Blue) color balls represents the Te(Se) atoms.

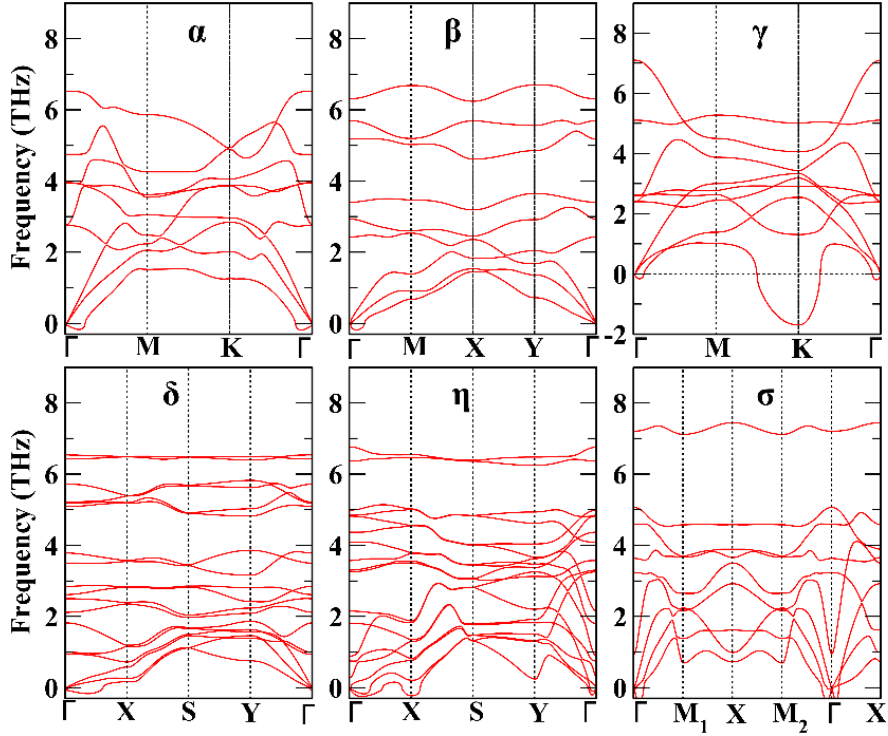

**Figure 2** Phonon dispersions of TeSe$_2$ allotropes along high symmetry lines.

### 3.2 Electronic Properties

The calculated electronic band structures for 2D TeSe$_2$ polytypes have been presented in figure 3 and respective band gaps have been tabulated in table 2. The available literature suggests 2D TeSe$_2$ as an indirect band gap semiconductor in α and β form [27, 30, 35]. For example, α-TeSe$_2$ has been reported as an indirect band gap semiconductor with band gap of 0.54 eV using GGA-PBE functional and 0.43 eV with PBE+SOC [35]. Our calculated value of 0.59 eV using GGA-PBE functional and 0.34 eV with PBE+SOC for α-TeSe$_2$ is in good agreement with earlier report [35]. Note that we find a new semiconducting allotrope of TeSe$_2$ (δ-TeSe$_2$) having direct band gap of 1.60 eV at high symmetry Γ point. However the difference between direct and indirect band gap is 5 *m*eV. Our calculations reveal that all the polytypes are semiconducting in nature with band gap ranging from 0.34 eV (η-TeSe$_2$) to 1.80 eV (σ-TeSe$_2$). The total density of states reveal zero density of states in the vicinity of Fermi level confirming the presence of finite energy gap in all six allotropes (figure 4). Also, a symmetric density of states for both the spins (spin up and spin down) gives a clear confirmation of nonmagnetic character of 2D-TeSe$_2$ polytypes. The magnitude of density of states is high for δ-TeSe$_2$ and η-TeSe$_2$ as number of atoms per unit cell for these phases are more as compared to other phases. It is noteworthy that the density of states show asymmetry in the vicinity of Fermi level as far as the valance band and conduction band is concerned indicating its good thermoelectric properties.

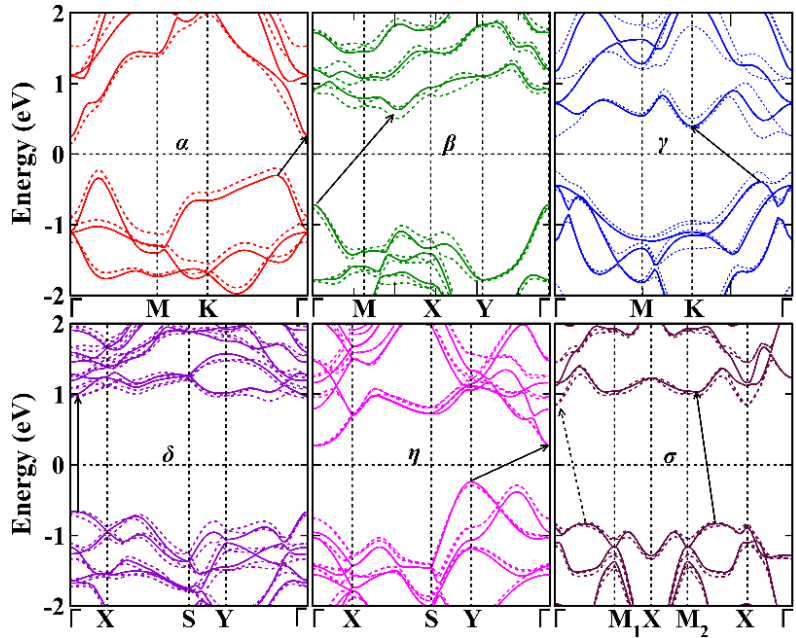

**Figure 3**. Electronic band structure of TeSe$_2$ allotropes calculated using PBE (solid lines) and PBE+SOC (dotted lines).

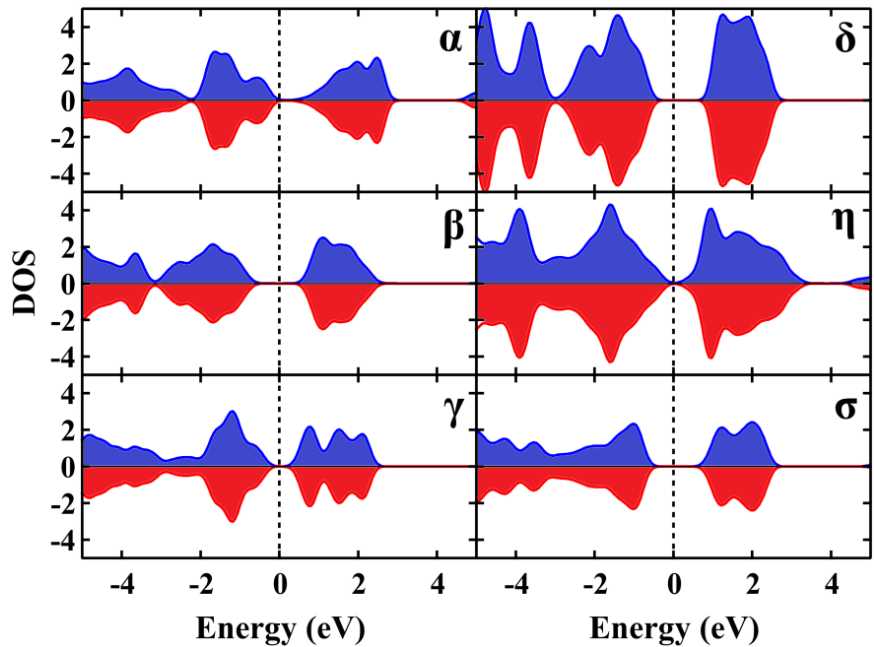

**Figure 4** Spin polarized total density of states for TeSe$_2$ allotropes. Blue and red color represents spin up and spin down density of states. Vertical dotted line represents the Fermi Level (set at 0eV).

Furthermore, work function is an important property of any surface which plays a significant role in device properties and performance. The Work function ($\Phi$) has been calculated for TeSe$_2$ allotropes as [37, 38]

$$\Phi = E_{vac} - E_F$$

where $E_{vac}$ and $E_F$ are vacuum and Fermi level respectively. The calculated values of work function have been listed in table 2. The allotropic TeSe$_2$ exhibit work function ranging between 4.19-5.11 eV. The work function of TeSe$_2$ allotropes are generally larger than tellurene monolayer [36] . The α- and γ-TeSe$_2$ have work function similar to MoS$_2$ (5.11 eV) [39]. The β and δ allotrope have work function smaller than that of graphene (4.51 eV)[40].

**Table 2** Electronic band gap (E$_g$) and work function ($\Phi$) for TeSe$_2$ allotropes.

| Allotrope | E$_g$ (eV) | | $\Phi$ (eV) |
|---|---|---|---|
| | PBE | PBE+SOC | |
| α | 0.59, 0.53[a], 0.54[b] | 0.34, 0.43[b] | 5.11 |
| β | 1.32 | 1.27 | 4.19 |
| γ | 0.79 | 0.62 | 5.11 |
| δ | 1.60 | 1.60 | 4.36 |
| η | 0.50 | 1.10 | 4.56 |
| σ | 1.80 | 1.64 | 4.56 |

[a]Ref [27], [b]Ref [35]

To understand the origin of valance band maxima (VBM) and conduction band minima (CBM) we have analyzed orbital projected density of states for *Te* and *Se* atoms (figure 5). For all allotropes VBM and CBM is dominated by *Se-4p* orbitals except for σ-TeSe$_2$. In σ-TeSe$_2$ VBM is dominated by *Te-5p* orbitals while CBM is *Se-4p* dominant suggesting the localization of charge carriers on different atoms of same 2D system. It indicates that this phase is suitable for solar cell applications by means of fabrication of appropriate Van-der-Waal hetero-structures where physical separation of charge carriers is required. A significant hybridization of *Te-p* and *Se-p* orbitals has been found in the vicinity of Fermi level for β-, δ-, and η-TeSe$_2$. This can be attributed to the structural differences among all the phases of TeSe$_2$.

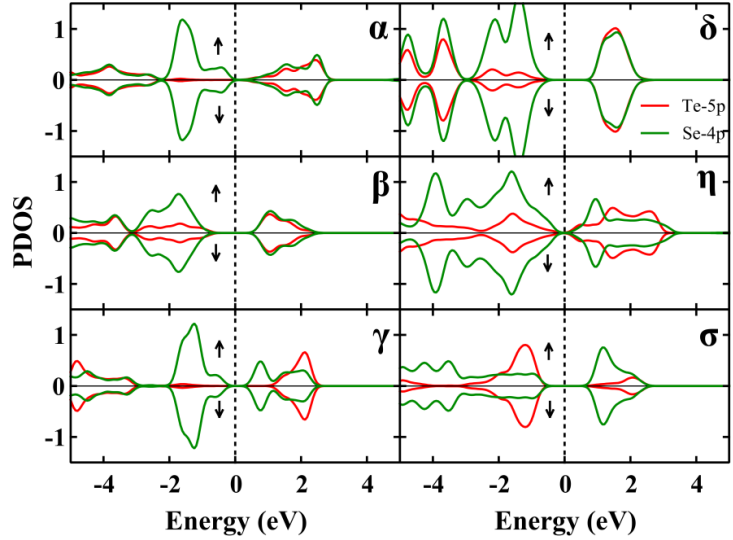

**Figure 5** Spin polarized orbital projected density of states for TeSe$_2$ allotropes. Arrows represents spin up and spin down density of states. Vertical dotted line represents the Fermi Level (set at 0 eV).

*3.2.1 STM topographic analysis and tunneling current characteristics*

Scanning tunneling microscopy (STM) is known for revealing structural and electronic properties. STM topographs obtained are presented in figure 6 (a-f). The STM topographical images have been obtained using Tersoff and Hamann formalism [41, 42] within the STM like setup as presented in figure 6(g). STM like topographs have been obtained at V= 1.0 eV between sample and tip. The *Se* atom can be expected to produce brighter spot in STM topographs due to its higher contribution to DOS as compared to *Te* atoms (figure 5). The α, γ, η and σ phase exhibit expected bright spots due to *Se* atoms and less bright spots due to *Te* atoms. However, geometrical effects dominance has been observed for β and δ allotrope having bright spot for atoms closest to STM tip. Both β and δ phase exhibit bright (dark) spots due to Te(Se) atom. Note that *Se* also show an intense bright and one less bright shade in STM topographs for δ-TeSe$_2$. Thus, different structural phases can be clearly distinguished with an STM images.

The bright spots observed in STM topographs indicate the contribution to the larger magnitude of tunneling current due to occupied energy states. In order to get an insight to distinguish the electronic structure, we now calculate the tunneling current characteristics using a model setup employed in the scanning tunneling microscope (STM) measurements. The tunneling current characteristics for TeSe$_2$ allotropes are presented in figure 6(h) between the bias ranges of -1.5 to +1.5 V. Here bias is said to be forward when sample is set at positive potential and electrons tunneling take place from tip to sample. All the allotropes are semiconducting as can be seen no current for the bias window of ± 0.2 V for α-, γ-, η-TeSe$_2$ and bias window of -0.2 to +1.0 V for β-, δ- and σ-TeSe$_2$. The magnitude of current is larger for α, η and γ allotropes while β, δ and η allotropes exhibit relatively lower magnitude of current. The magnitude of current in forward (reverse) bias is attributed to availability of number of unoccupied (occupied) energy states. The difference in characteristics can be understood in terms of density of states. For example, for a given bias, large magnitude of current can be seen for η as compared to γ allotrope. This difference is attributed to the convolution of DOS of *Se* atoms in γ allotrope and both *Te*, *Se* DOS in η allotrope with tip density of states.

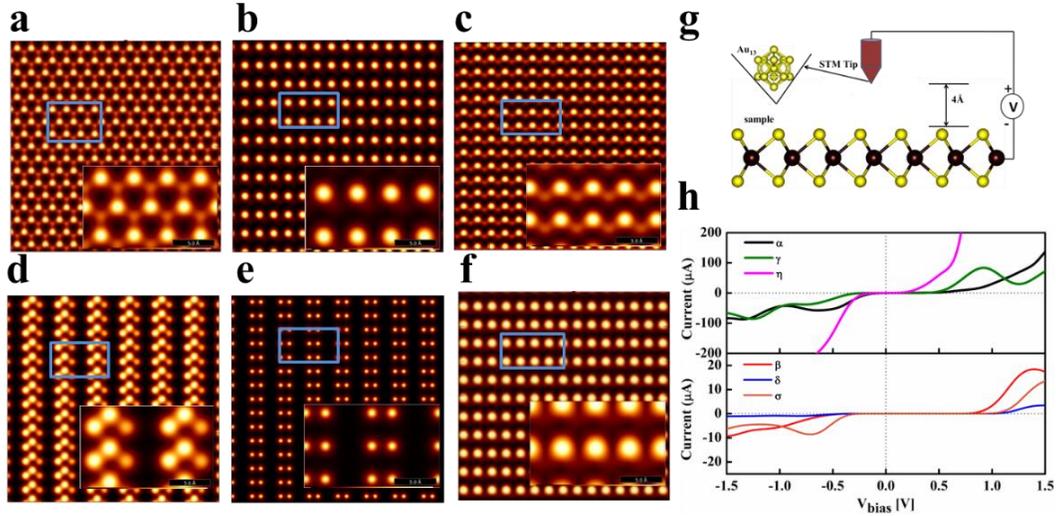

**Figure 6** STM topographs (a to f) at V= 1.0 V, schematic of STM model (g) and tunneling current characteristics for TeSe$_2$ allotropes (h).

### 3.3 Thermoelectric Properties

#### *3.3.1 Temperature dependent thermoelectric properties*

Now, we discuss the temperature dependent thermoelectric properties of TeSe$_2$ allotropes. A super cell approach has been used to obtain interatomic force constants using SIESTA [31] and lattice thermal conductivities were obtained using GOLLUM code [34]. The calculated *G*, *S*, *PF*, *k* and *ZT* as a function of temperature are presented in figure 7. A large value of *S* can be noticed for lower temperatures which decreases with increase in temperature for all the allotropes. At room temperature, value of *S* is maximum (3.06 mV/K) for delta allotrope and minimum (1.11 mV/K) for alpha allotrope. Note that Seebeck coefficient has negative Seebeck coefficient in entire temperature range for η-TeSe$_2$ indicating holes as majority carrier. On the other hand, G show inverse behavior where highest value of 0.8 μG$_0$ has been found for alpha allotrope at room temperature. A significant increase in G begins above 600 K indicating metallic transport behavior above this temperature.

The higher value of G increases the PF for alpha phase. The power factor reaches to 150 pW/K for α-TeSe$_2$ at 1000 K which is significantly higher than any other phase. A significant increase in power factor for alpha allotrope is attributed to the increase in electrical conductivity (G). The cumulative thermal conductivity ($k_{ph}+k_e$) has been found to depend weakly on temperature below 600 K and maintains a phase dependent characteristic value which is generally less than 61 pW/K. Note that cumulative thermal conductivity increases beyond 600K for gamma phase. The increase in *PF* and lower value of cumulative thermal conductivity results an increase in *ZT* for α-TeSe$_2$ and minimum for β-TeSe$_2$. The higher figure of merit lies in higher temperature range which indicate their importance for high temperature thermoelectric applications such as power generation.

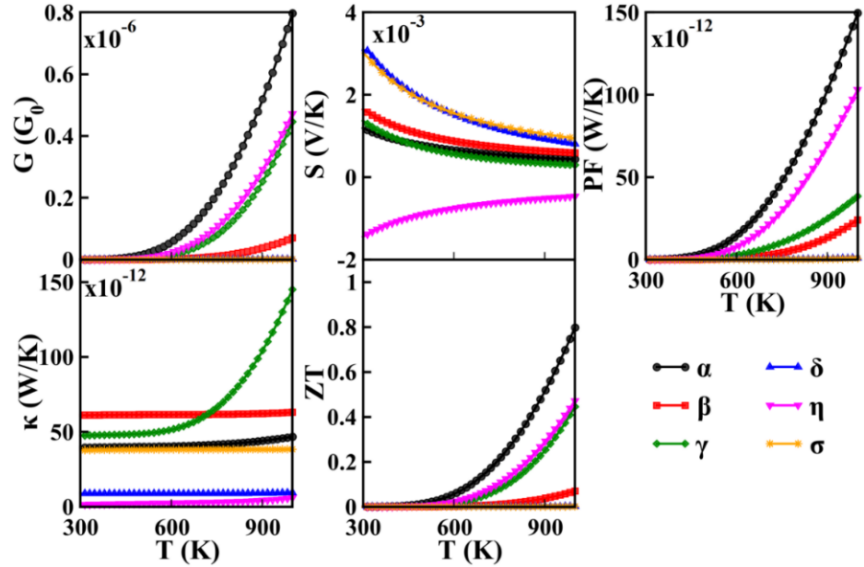

**Figure 7** Calculated temperature dependent electrical conductivity (G), seebeck coefficient (S), power factor (PF), cumulative thermal conductivity ($\kappa = k_{ph}+k_e$) and ZT for 2D-TeSe$_2$ allotropes.

### 3.3.2 Chemical Potential tunable thermoelectric properties.

Furthermore, we study the effect of chemical potential on ZT of TeSe$_2$ allotropes at 300 K. Figure 8 show G, S, PF and ZT as a function of chemical potential (μ), where μ is positive(negative) for n-type (p-type) doping. The change in chemical potential toward the conduction (valence) increases the electron (hole) carriers in the TeSe$_2$ and thereby increases electrical conductivity.

The electrical conductivity is more for p-type doping as compared to n-type doping. An enhancement in G has been found for γ-TeSe$_2$ with *p*-type doping while for all other cases it modulate continuously with chemical potential. An observable peak in electrical conductivity near 0.5 eV for *n*-type doping indicate the presence of mid-gap states in η- and γ-TeSe$_2$. The Seebeck coefficients of TeSe$_2$ for all cases exhibit two peaks around the Fermi level (μ = 0), because the Fermi level is close to the band edges where the energy contributions are higher. The sign of Seebeck coefficient changes from positive to negative at charge neutrality point (μ = 0) corresponding to the change in the type of majority charge carriers. The room temperature Seebeck coefficient is maximum of ~ 4 V/K for δ-TeSe$_2$ which is attributed to asymmetric *Se-p* contributions around Fermi level. The *G* and *S* defines the power factor (*PF*) which play an important role in maximizing *ZT*.

The larger values of Seebeck coefficient around Fermi level and very small electrical conductivity of TeSe$_2$ results in smaller PF in this region. The modulation in chemical potential within the 1 eV window modulates the PF significantly and hence ZT. The ZT value of delta phase reaches to ~3.1 for p-type doping and 2.3 for n-type doping. It is noticeable that the higher value is achievable at a critical value of chemical potential. The highest value of ZT for delta allotrope is attributed to its relatively high Seebeck coefficient and low thermal conductivity.

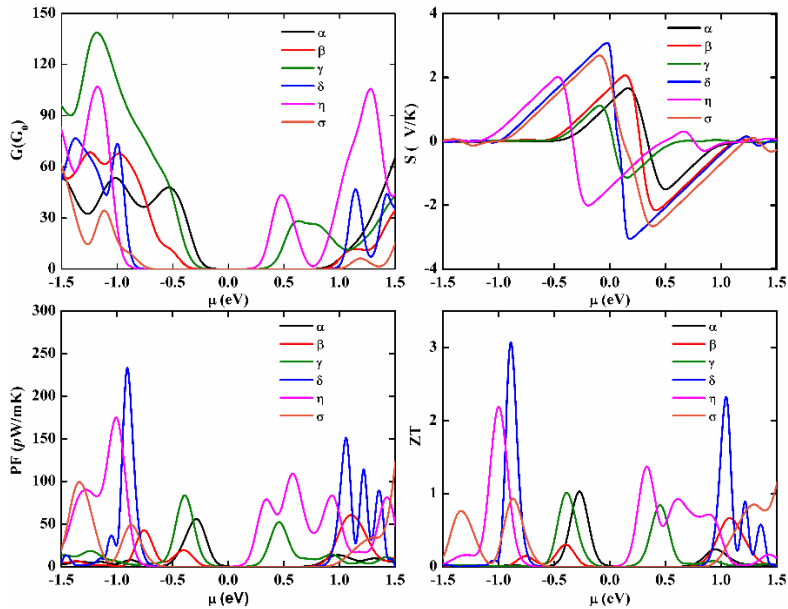

**Figure 8** Calculated electrical conductivity (*G*), Seebeck coefficient (*S*), power factor (*PF*) and *ZT* as a function of chemical potential for TeSe$_2$ allotropes at 300K.

## 4. Conclusions

In this study, we investigate relative energetic and kinetic stability of six monolayer $TeSe_2$ allotropes. The difference in relative formation energy are less than 5 meV for β-, δ-, η- and σ-$TeSe_2$ which demonstrate their possible coexistence in experimental realization. The γ phase has been demonstrated energetically and kinetically instable. The band structure calculations reveal all six allotropes semiconducting with band gap lying in IR region for α-, β-, γ-, η-$TeSe_2$ and Visible region for δ-, σ-$TeSe_2$. The structural asymmetry in $TeSe_2$ has been clearly captured in simulated scanning tunnel microscopy. The η phase exhibit large magnitude of tunneling current as compared to γ phase due to dominance of both Te-p and Se-p density of states. The thermoelectric efficiency (*ZT*) can be maximized up to 3.1 with p-type doping in δ-$TeSe_2$. Our computational study affirms that the thermoelectric figure of merit (*ZT*) in $TeSe_2$ monolayers can be tailored via temperature and chemical potential engineering. Therefore, our present study is compelling from an experimental perspective.

## Conflict of Interest

There is no conflict of interest to declare.

## Acknowledgements

Helpful discussions with Ritika Rani and Kuldeep Kumar are highly acknowledged.

# References

1. Geim, A.K., *Graphene: status and prospects.* science, 2009. **324**(5934): p. 1530-1534.
2. Shunhong, Z., et al., *Penta-graphene: A new carbon allotrope.* Proceedings of the National Academy of Sciences, 2015. **112**(8).
3. Sheng, X.-L., et al., *T-carbon: a novel carbon allotrope.* Physical review letters, 2011. **106**(15): p. 155703.
4. Manzeli, S., et al., *2D transition metal dichalcogenides.* Nature Reviews Materials, 2017. **2**(8): p. 17033.
5. Chen, Y., et al., *In-Plane Anisotropic Thermal Conductivity of Few-Layered Transition Metal Dichalcogenide Td-WTe2.* Advanced Materials, 2019. **31**(7): p. 1804979.
6. Sharma, M., A. Kumar, and P. Ahluwalia, *Electron transport and thermoelectric performance of defected monolayer MoS2.* Physica E: Low-dimensional Systems and Nanostructures, 2019. **107**: p. 117-123.
7. Li, W., et al., *Gapless MoS 2 allotrope possessing both massless Dirac and heavy fermions.* Physical Review B, 2014. **89**(20): p. 205402.
8. Xu, Y., Z. Gan, and S.-C. Zhang, *Enhanced thermoelectric performance and anomalous Seebeck effects in topological insulators.* Physical review letters, 2014. **112**(22): p. 226801.
9. Zhang, D.-C., et al., *Thermoelectric properties of β-As, Sb and Bi monolayers.* RSC advances, 2017. **7**(39): p. 24537-24546.
10. Morales-Ferreiro, J., et al., *First-principles calculations of thermoelectric properties of IV–VI chalcogenides 2D materials.* Frontiers in Mechanical Engineering, 2017. **3**: p. 15.
11. Dutta, M., T. Ghosh, and K. Biswas, *Electronic structure modulation strategies in high-performance thermoelectrics.* APL Materials, 2020. **8**(4): p. 040910.
12. Guo, S.-D., *Spin-orbit and strain effect on power factor in monolayer MoS2.* Computational Materials Science, 2016. **123**: p. 8-13.
13. Dresselhaus, M., et al., *Low-dimensional thermoelectric materials.* Physics of the Solid State, 1999. **41**(5): p. 679-682.
14. Hicks, L. and M.S. Dresselhaus, *Effect of quantum-well structures on the thermoelectric figure of merit.* Physical Review B, 1993. **47**(19): p. 12727.
15. Feng, B., et al., *Experimental realization of two-dimensional boron sheets.* Nature chemistry, 2016. **8**(6): p. 563-568.
16. Zhu, F.-f., et al., *Epitaxial growth of two-dimensional stanene.* Nature materials, 2015. **14**(10): p. 1020-1025.
17. Li, L., et al., *Black phosphorus field-effect transistors.* Nature nanotechnology, 2014. **9**(5): p. 372.
18. Zhu, Z., et al., *Multivalency-driven formation of Te-based monolayer materials: a combined first-principles and experimental study.* Physical review letters, 2017. **119**(10): p. 106101.
19. Chen, J., et al., *Ultrathin β-tellurium layers grown on highly oriented pyrolytic graphite by molecular-beam epitaxy.* Nanoscale, 2017. **9**(41): p. 15945-15948.
20. Wu, L., et al., *2D tellurium based high-performance all-optical nonlinear photonic devices.* Advanced Functional Materials, 2019. **29**(4): p. 1806346.
21. Gao, Z., G. Liu, and J. Ren, *High thermoelectric performance in two-dimensional tellurium: An ab initio study.* ACS applied materials & interfaces, 2018. **10**(47): p. 40702-40709.
22. Shi, Z., et al., *Two-Dimensional Tellurium: Progress, Challenges, and Prospects.* Nano-Micro Letters, 2020. **12**: p. 1-34.
23. Apte, A., et al., *Polytypism in ultrathin tellurium.* 2D Materials, 2018. **6**(1): p. 015013.
24. Wang, C., et al., *Charge-governed phase manipulation of few-layer tellurium.* Nanoscale, 2018. **10**(47): p. 22263-22269.
25. Wang, J., et al., *Electric Field-Tunable Structural Phase Transitions in Monolayer Tellurium.* ACS omega, 2020. **5**(29): p. 18213-18217.

26. Liu, G., H. Wang, and G.-L. Li, *Thermal expansion and vibrational properties of α-Se and α-TeSe2 based on first-principles calculations.* Solid State Communications, 2020: p. 113912.
27. Naseri, M., et al., *Density functional theory based prediction of a new two-dimensional TeSe2 semiconductor: A case study on the electronic properties.* Chemical Physics Letters, 2018. **707**: p. 160-164.
28. Chen, Y., et al., *Symmetry-breaking induced large piezoelectricity in Janus tellurene materials.* Physical Chemistry Chemical Physics, 2019. **21**(3): p. 1207-1216.
29. Min, J., et al., *Tunable visible-light excitonic absorption and high photoconversion efficiency in two-dimensional group-VI monolayer materials.* Physical Review B, 2019. **100**(8): p. 085402.
30. Wu, B., et al., *A new two-dimensional TeSe 2 semiconductor: indirect to direct band-gap transitions.* Science China Materials, 2017. **60**(8): p. 747-754.
31. Ordejón, P., E. Artacho, and J.M. Soler, *Self-consistent order-N density-functional calculations for very large systems.* Physical Review B, 1996. **53**(16): p. R10441.
32. Troullier, N. and J.L. Martins, *Efficient pseudopotentials for plane-wave calculations.* Physical review B, 1991. **43**(3): p. 1993.
33. Togo, A. and I. Tanaka, *First principles phonon calculations in materials science.* Scripta Materialia, 2015. **108**: p. 1-5.
34. Ferrer, J., et al., *GOLLUM: a next-generation simulation tool for electron, thermal and spin transport.* New Journal of Physics, 2014. **16**(9): p. 093029.
35. Debela, T.T. and H.S. Kang, *Phase polymorphism and electronic structures of TeSe 2.* Journal of Materials Chemistry C, 2018. **6**(38): p. 10218-10225.
36. Rani, R., M. Sharma, and R. Sharma, *Optical Anisotropy in Tellurene and its Janus Allotropes--A first principle Study.* arXiv preprint arXiv:2007.13066, 2020.
37. Shan, B. and K. Cho, *First principles study of work functions of single wall carbon nanotubes.* Physical review letters, 2005. **94**(23): p. 236602.
38. Shan, B. and K. Cho, *First-principles study of work functions of double-wall carbon nanotubes.* Physical Review B, 2006. **73**(8): p. 081401.
39. Guo, M., et al., *Edge dominated electronic properties of MoS2/graphene hybrid 2D materials: edge state, electron coupling and work function.* Journal of Materials Chemistry C, 2017. **5**(20): p. 4845-4851.
40. Liang, S.-J. and L. Ang, *Electron thermionic emission from graphene and a thermionic energy converter.* Physical Review Applied, 2015. **3**(1): p. 014002.
41. Tersoff, J. and D. Hamann, *Theory and application for the scanning tunneling microscope.* Physical review letters, 1983. **50**(25): p. 1998.
42. Feenstra, R.á., J.A. Stroscio, and A.á. Fein, *Tunneling spectroscopy of the Si (111) 2× 1 surface.* Surface science, 1987. **181**(1-2): p. 295-306.